\documentclass[aps,prl,twocolumn,superscriptaddress,showkeys]{revtex4-2}  
\usepackage{graphicx}  
\usepackage{dcolumn}   
\usepackage{bm}        
\usepackage{amssymb}   
\usepackage{epsfig}

\begin{document}

\title{Effect of the sample work function on alkali metal dosing induced electronic structure change}

\author{Saegyeol Jung}
\affiliation{Center for Correlated Electron Systems, Institute for Basic Science, Seoul, 08826, Korea}
\affiliation{Department of Physics and Astronomy, Seoul National University, Seoul, 08826, Korea}

\author{Yukiaki Ishida}
\affiliation{Center for Correlated Electron Systems, Institute for Basic Science, Seoul, 08826, Korea}
\affiliation{Institute for Solid State Physics (ISSP), The University of Tokyo, Kashiwa, Chiba 277-8581, Japan}

\author{Minsoo Kim}
\affiliation{Center for Correlated Electron Systems, Institute for Basic Science, Seoul, 08826, Korea}
\affiliation{Department of Physics and Astronomy, Seoul National University, Seoul, 08826, Korea}

\author{Masamichi Nakajima}
\affiliation{Department of Physics, Osaka University, Toyonaka, Osaka 560-0043, Japan}

\author{Shigeyuki Ishida}
\affiliation{National Institute of Advanced Industrial Science and Technology (AIST), Tsukuba, Ibaraki 305-8568, Japan}

\author{Hiroshi Eisaki}
\affiliation{National Institute of Advanced Industrial Science and Technology (AIST), Tsukuba, Ibaraki 305-8568, Japan}

\author{Woojae Choi}
\affiliation{Department of Emerging Materials Science, DGIST, Daegu, 711-873, Korea}

\author{Yong Seung Kwon}
\affiliation{Department of Emerging Materials Science, DGIST, Daegu, 711-873, Korea}

\author{Jonathan Denlinger}
\affiliation{Advanced Light Source, Lawrence Berkeley National Laboratory, California 94720, USA}

\author{Toshio Otsu}
\affiliation{Institute for Solid State Physics (ISSP), The University of Tokyo, Kashiwa, Chiba 277-8581, Japan}

\author{Yohei Kobayashi}
\affiliation{Institute for Solid State Physics (ISSP), The University of Tokyo, Kashiwa, Chiba 277-8581, Japan}

\author{Soonsang Huh}\email{Corresponding autor: sshuhss@gmail.com}
\affiliation{Center for Correlated Electron Systems, Institute for Basic Science, Seoul, 08826, Korea}
\affiliation{Department of Physics and Astronomy, Seoul National University, Seoul, 08826, Korea}

\author{Changyoung Kim}\email{Corresponding autor: changyoung@snu.ac.kr}
\affiliation{Center for Correlated Electron Systems, Institute for Basic Science, Seoul, 08826, Korea}
\affiliation{Department of Physics and Astronomy, Seoul National University, Seoul, 08826, Korea}

\begin{abstract}
	\vspace{1\baselineskip}
Alkali metal dosing (AMD) has been widely used as a way to control doping without chemical substitution. This technique, in combination with angle resolved photoemission spectroscopy (ARPES), often provides  an opportunity to observe unexpected phenomena. However, the amount of transferred charge and the corresponding change in the electronic structure vary significantly depending on the material. Here, we report study on the correlation between the sample work function and alkali metal induced electronic structure change for three iron-based superconductors: FeSe, Ba(Fe$_{0.94}$Co$_{0.06}$)$_{2}$As$_{2}$ and NaFeAs which share a similar Fermi surface topology. Electronic structure change upon monolayer of alkali metal dosing and the sample work function were measured by ARPES. Our results show that the degree of electronic structure change is proportional to the difference between the work function of the sample and Mulliken’s absolute electronegativity of the dosed alkali metal. This finding provides a possible way to estimate the AMD induced electronic structure change.
\vspace{2\baselineskip}
\end{abstract}


\maketitle

\section{1. Introduction}
Alkali metal dosing (AMD) on a material often leads to transfer of charge from alkali metal to the surface of the material~\cite{muscat1986coverage,aruga1989alkali,diehl1997current,politano2008nature}. Due to such property, AMD became an important method to dope materials with electrons, and studies using AMD have been performed on various materials, such as photoemissive materials~\cite{sommer1968photoemissive,mine1994photoemissive}, organic charge transport materials~\cite{haddon1991conducting,ganzorig2001alkali,parthasarathy2001lithium} and catalysts~\cite{campbell1982surface,heskett1985photoemission,lackey1987chemical}. In recent years, AMD has been extensively used in angle resolved photoemission spectroscopy (ARPES) studies of correlated materials, in which electronic structure change induced by surface electron doping from AMD is investigated. This new tuning parameter in ARPES has been exploited in the observation of novel phenomena~\cite{zhang2014direct,kim2015observation,miyata2015high,ohta2006controlling}. However, it was found in numerous studies that the extent of the electronic structure changes upon monolayer of alkali metal dosing varies from system to system~\cite{kyung2016enhanced,seo2016superconductivity} and the exact cause for the variation is not well understood as there has not been systematic studies to investigate the cause for the variation.

In previous AMD studies, it was reported that the work function modulation upon AMD is proportional to the difference between the work function of the sample and the Mulliken’s absolute electronegativity of the alkali metal~\cite{alton1986semi}. Since it was theoretically suggested that the amount of charge transfer upon AMD is proportional to this work function modulation ~\cite{leung2003relationship}, it is reasonable to assume that the electronic structure change induced by charge transfer upon AMD is related to the difference between the work function of the sample and the electronegativity of the dosed alkali metal. In that respect, it is important to investigate how the degree of the electronic structure change upon AMD is related to the difference between the work function of the sample and the electronegativity of the alkali metal.

  In this article, we report work function measurements of three iron-based superconductors (FeSe, Ba(Fe$_{0.94}$Co$_{0.06}$)$_{2}$As$_{2}$ and NaFeAs) with laser-based ARPES. Electronic structure change upon AMD is also investigated. It is revealed that, for materials with a similar Fermi surface topology, the electronic structure change upon monolayer of alkali metal dosing has linear correlation with the difference between the sample work function and the electronegativity of the alkali metal. Our results provide certain basis for future ARPES studies with alkali metal dosing.

\begin{figure}
\centering \epsfxsize=6.6cm \epsfbox{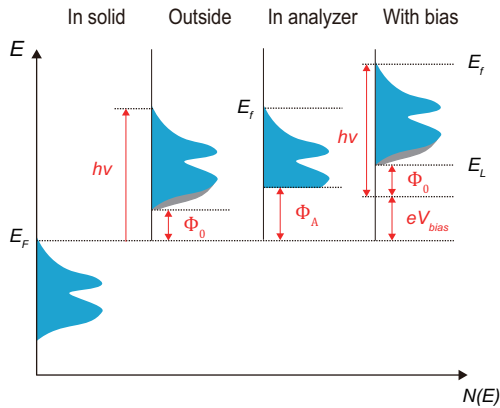} \caption{(Color online). Schematic energy diagrams for the photoemission process and work function measurement. The occupied part of the density of states (leftmost), distribution of photoelectrons kicked out from the sample (second from left), distribution of photoelectrons in the electron analyzer (third) and the same distribution but with a bias voltage $V_{bias}$ (rightmost). $E_{f}$, $E_L$, $\Phi_{0}$, $\Phi_{A}$ denote level of Fermi edge , measured lowest kinetic energy of photoelectrons at low-energy cutoff, work function of the sample and work function of electron analyzer, respectively. The area shaded in gray represents the background from inelastically scattered electrons. }
\label{fig1}
\end{figure}

\section{2. Experimental method}
Work function measurements were performed with a 5.988 eV fiber-laser source~\cite{ishida2016high} with a lab-based system at Seoul National University~\cite{Ishida2020workfunction}. The energy resolution (convolution of the analyzer's resolution and the bandwith of the light source) acquired from polycrystalline gold with no bias voltage is about 12 meV. In measuring $E_{L}$, which refers to the cutoff on the slowest end of the spectrum,  we applied -3.232, -6.477 and -6.353 V bias voltages to FeSe, Ba(Fe$_{0.94}$Co$_{0.06}$)$_{2}$As$_{2}$ and NaFeAs, respectively, with commercial dry cell batteries. The accuracy in the applied bias voltage is about 0.01\% in each measurement. ARPES measurements with alkali metal dosing were performed at the beam line 4.0.3 of the Advanced Light Source. Rb was chosen for alkali metal dosing because of its similar electronegativity value compared to the work function of NaFeAs. The coverage of Rb was determined from Rb core level spectra, as was done in previous studies~\cite{kyung2016enhanced}. To avoid surface contamination effects, work function was measured within 8 hours after cleave at a pressure better than $6\times10^{-11}$ Torr.

\section{3. Results and discussion}
Fig. 1 shows schematic energy diagrams for the photoemission process and work function measurements. The first panel shows the occupied part of the density of states $(N(E))$ in the sample. When light with an energy of $hv$ impinges on the sample, electrons are photo-emitted as illustrated in the second panel. The level of Fermi edge is now located $hv$ above the Fermi level ($E_{F}$). In addition to the photoelectrons, we also have the background from inelastically scattered electrons as depicted by the grey shaded area. The distribution is cut by the vacuum level, resulting in a low-energy cutoff which represents the work function. When the sample work function is smaller than that of the electron analyzer, this low energy cutoff cannot be observed as illustrated in the third panel. Thus, the sample is usually negatively biased to avoid the problem with this high analyzer work function~\cite{Pfau2020lowworkfunction}, as illustrated in the fourth panel. From the level of low-energy cutoff ($E_{L}$) and Fermi edge ($E_{f}$), the work function of the sample $\Phi_{0}$ can be obtained by~\cite{CardonaLey}

\begin{equation}
\Phi_{0}=hv-(E_{f}-E_{L})
\end{equation}
where $hv$ is the photon energy of the light source.

\begin{figure}
	\centering \epsfxsize=8.0cm \epsfbox{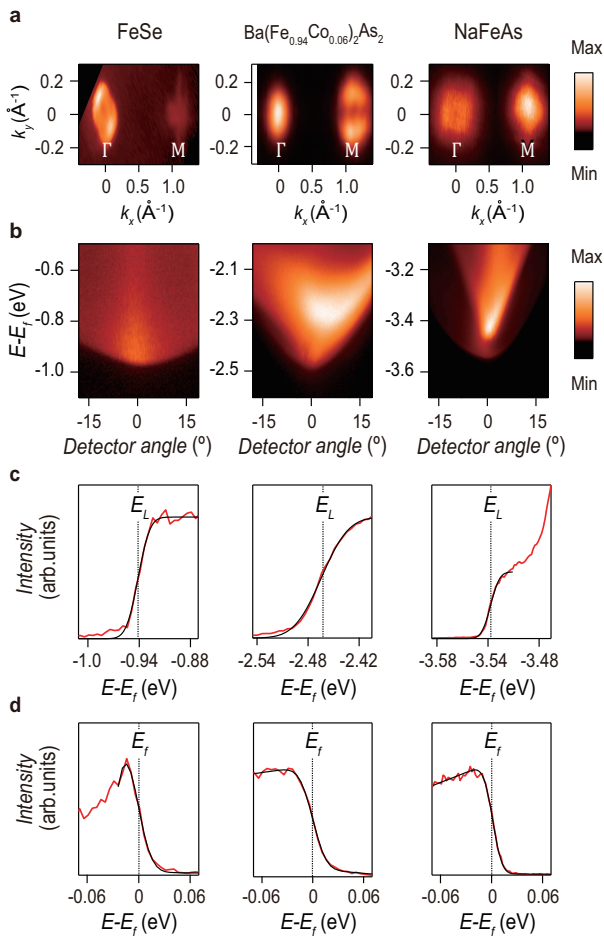} \caption{(Color online). (a) ARPES Intensity maps of FeSe, Ba(Fe$_{0.94}$Co$_{0.06}$)$_{2}$As$_{2}$ and NaFeAs at the level of fermi edge. (b) ARPES spectra of low-energy cutoff and (c) corresponding normal emission EDCs of same materials. Here, the EDCs were integrated over $\pm 0.2^{\circ}$ around the vertex of parabola . (d) EDCs near the level of fermi edge measured at $\Gamma$. Black solid lines in (c) and (d) represent fitting result with gaussian convoluted step function and Fermi-dirac distribution function, respectively. }
	\label{fig2} 
\end{figure}

Fig. 2(a) shows Fermi surface maps of three iron-based superconductors: FeSe, Ba(Fe$_{0.94}$Co$_{0.06}$)$_{2}$As$_{2}$ and NaFeAs. It clearly shows a similar Fermi surface topology for these three superconductors. Since the extent of electronic structure change induced by AMD is set by the density of states near the level of fermi edge, this similarity in the Fermi surface topology is important in finding the correlation between work function and electronic structure change induced by AMD. Fig. 2(b) shows low energy side of ARPES spectra with bias voltage for same materials. Note that all the spectra have parabolic low-energy cutoff for which the parabolic shape comes from the non-zero in-plane momentum of photoelectrons at the cutoff~\cite{Ishida2020workfunction}. As the $E_{L}$ in equation (1) should be measured at the vertex of the parabola, the energy distribution curve (EDC) at the vertex position was used to determine the $E_{L}$ (see Fig. 2(c)). $E_{f}$ in equation (1) was obtained from EDCs near the level of fermi edge at $\Gamma$ (see Fig. 2(d)). $E_{L}$ and $E_{f}$ were precisely determined by fitting each EDC with step function and Fermi-dirac distribution function convoluted with gaussian distribution, respectively. We note that EDC near low-energy cutoff was fitted with gaussian convoluted step function because the width of the low-energy cutoff is not blurred by the bandwidth of the light nor by the Fermi-Dirac distribution function. From these level of low-energy cutoff $E_{L}$ and fermi edge  $E_{f}$ of the sample, the work function is calculated using equation (1). The obtained work function values ($\Phi_{0}$), gaussian width of step function near $E_{L} $ ($\gamma_{L}$) and gaussian width of Fermi-dirac distribution near $E_{f} $ ($\gamma_{f}$) are summarized in Table 1 .

\begin{table}
	{\footnotesize
	\centering
	\renewcommand{\arraystretch}{1.5}
	\renewcommand{\tabcolsep}{2.0mm}
	\begin{tabular}{c|ccc||c|c}
		\noalign{\smallskip}\noalign{\smallskip}\hline\hline
	
			&  $\Phi_{0}$ & $\gamma_{L}$   & $\gamma_{f}$   &   & $\chi$  \\
		\hline
		
			FeSe   & 5.047 & 0.025   & 0.022      & Na & 2.843  \\
		
	Ba(Fe$_{0.94}$Co$_{0.06}$)$_{2}$As$_{2}$  & 3.524 & 0.058  & 0.025     & K  & 2.421 \\
	
			NaFeAs & 2.451 & 0.02 & 0.019         & Rb & 2.331 \\
		\hline
		\hline

	\end{tabular}
}
\caption{ Work function $\Phi_{0}$ of three iron-based superconductors : FeSe, Ba(Fe$_{0.94}$Co$_{0.06}$)$_{2}$As$_{2}$ and NaFeAs. $\gamma_{L}$ and $\gamma_{f}$ indicate gaussian width of step function near $E_{L} $ and Fermi-dirac distribution near $E_{f}$, respectively. 
	$\gamma_{L}$ includes measurement accuracy in bias voltage. We added Mullikan's absolute electronegativity $\chi=\frac{1	}{2}(I_{a}+E_{a})$ of alkali metals, where $I_{a}$ obtained from \cite{David} and $E_{a}$ obtained from \cite{Hotop}, respectively . All the values in Table 1 has unit of eV.}
\label{Table 1}
\end{table}

In previous ARPES studies with AMD~\cite{seo2016superconductivity,kyung2016enhanced}, it has been reported that the electron band at the M point shifts downward about 45 and 14 meV for Na monolayer dosed FeSe and K monolayer dosed Ba(Fe$_{0.94}$Co$_{0.06}$)$_{2}$As$_{2}$, respectively. Fig. 3(a) and (b) show previously reported ARPES spectra data and EDCs at M point in those studies~\cite{seo2016superconductivity,kyung2016enhanced}. Blue and red dashed lines in Fig. 3(a) and (b) represent position of the electron band at M point obtained from second derivative data (Fig. 3(d)) and its EDC (Fig. 3(e)). These results can be understood from larger difference between FeSe work function and Na electronegativity than that of K dosed Ba(Fe$_{0.94}$Co$_{0.06}$)$_{2}$As$_{2}$~\cite{David,Hotop} (see Table 1). Meanwhile, the work function of NaFeAs is found to be similar to the electronegativity of Rb. Therefore, it is reasonable to speculate that there would be very small band shift in Rb  dosed NaFeAs in comparison to the cases of Na dosed FeSe and K dosed Ba(Fe$_{0.94}$Co$_{0.06}$)$_{2}$As$_{2}$.

In order to confirm our speculation, we performed Rb dosing experiment on NaFeAs. ARPES spectra and EDCs at the M point of pristine (left) and Rb monolayer dosed (right) NaFeAs are shown in Fig. 3(c). The coverage of Rb was determined following the method reported in a previous study~\cite{kyung2016enhanced}. To determine the band shift more clearly, second derivative data and its EDC (Fig. 3(f)) are shown. Blue and red arrows in Fig. 3(f) indicate the peak position of pristine and Rb dosed samples. The result shows an upward band shift of about 3 meV after Rb dosing. This result shows that the amount of band shift is much smaller than that of FeSe and Ba(Fe$_{0.94}$Co$_{0.06}$)$_{2}$As$_{2}$ as we suspected.

\begin{figure*}[ht] 
	\centering \epsfxsize=17.5 cm \epsfbox{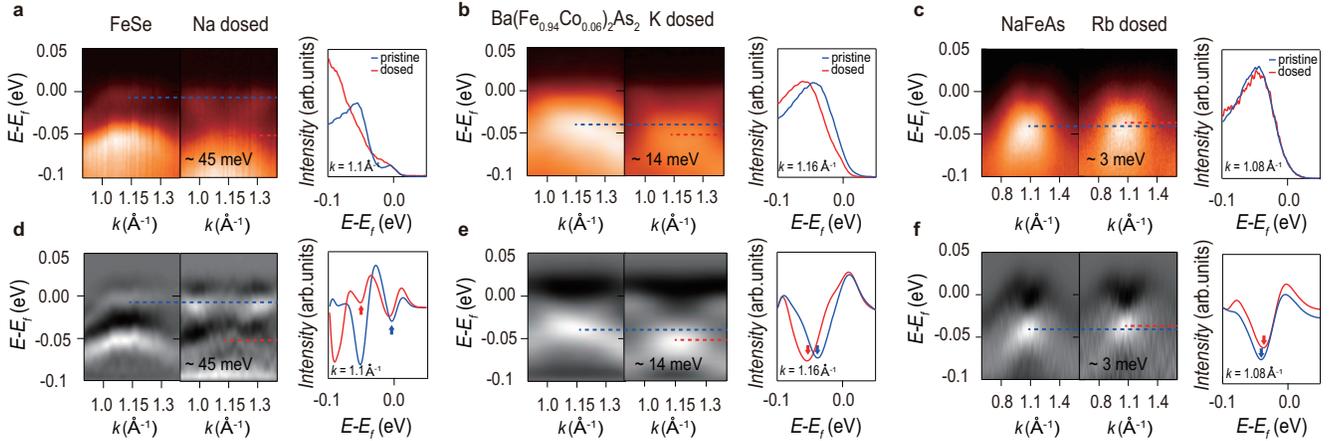} \caption{(Color online). ARPES spectra data and EDCs near the M point of (a) Na dosed FeSe, (b) K dosed Ba(Fe$_{0.94}$Co$_{0.06}$)$_{2}$As$_{2}$, adopted from ~\cite{kyung2016enhanced,seo2016superconductivity} and (c) Rb dosed NaFeAs. Blue and red dashed lines show position of the electron band at the M point obtained from pristine and dosed samples. 
		Second derivative ARPES spectra around the M point and  its EDCs at the M point of (d) Na dosed FeSe, (e) K dosed Ba(Fe$_{0.94}$Co$_{0.06}$)$_{2}$As$_{2}$ and (f) Rb dosed NaFeAs. Blue and red arrows mark peak positions of the pristine and dosed samples, respectively. The band shift downward about 45 and 14 meV for Na dosed FeSe and K dosed Ba(Fe$_{0.94}$Co$_{0.06}$)$_{2}$As$_{2}$, respectively. Meanwhile, the band shifts upward by 3 meV after Rb dosing. All the EDCs from the dosed sample are multiplied by 1.5-3 for an easier comparison.}
	\label{fig3}
\end{figure*}

From our result of work function measurement and dosing studies~\cite{kyung2016enhanced,seo2016superconductivity,ye2015simultaneous}, we obtained and plot the change in the electron band bottom position as a function of the difference between work function and electronegativity in Fig. 4. It is clear that there is a positive correlation between the two values; the band shift becomes larger as the work function increases. In other words, the change in the electronic structure is related to the difference between the work function of the sample and the electronegativity of the alkali metal.
A noticeable point is that the plot in Fig. 4 shows a weak linear relationship described by

\begin{equation}
\Delta E^{M}  = a(\Phi_{0} -\chi) - b
\end{equation}
where $\Delta E^{M}$ is the shift of the band at the M point, $\Phi_{0}$ and $\chi$ are the sample work function and the Mulliken's absolute electronegativity of dosed alkali metal. Here, a = 0.021  and b = 6.243 meV are the constants obatained from linear fitting. Coefficient of determination $R^{2}=0.9715$ was obtained from simple linear regression analysis.

This linear relation may be interpreted within a phenomenological model that used to describe the band shift in physisorbed graphene~\cite{khomyakov}. To apply this model on our case, we replace the density of states of graphene to the nearly free electron density of state in 2D since AMD generates a quasi-2D electron gas confined in few top layers. From this replacement, the phenomenological model shows the linear relation between the band shift and the difference between work function of sample and electronegativity of dosed alkali metal. Since these three iron-based superconductors share a similar Fermi surface topology, the linear relation in our data also provides an experimental evidence for the theory which expects linear relation between surface electron doping and work function modulation induced by AMD~\cite{leung2003relationship}.

 A possible reason for the slight deviation from linear behavior is that the electronic structure change induced by AMD cannot be completely described within a simple rigid band shift picture. Other factors such as localization of the alkali metal~\cite{kim2017microscopic}, additional electric field by alkali metal layer and band dependent doping effect~\cite{ito2018alkali} may also have some effects. Even though these factors should be considered for a concrete understanding, our results should be useful for the future ARPES studies using AMD.

\begin{figure}
	\centering \epsfxsize=8.6cm \epsfbox{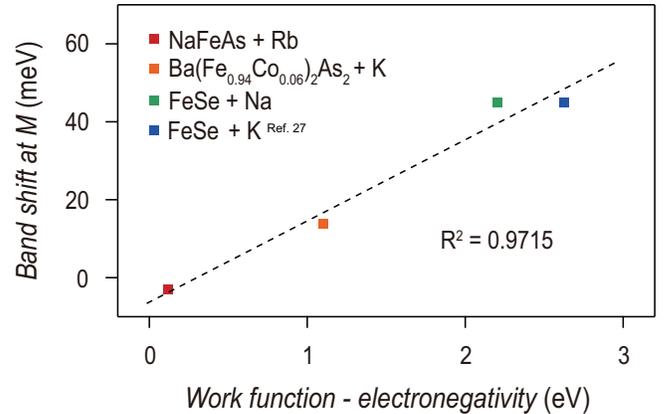} \caption{(Color online). Band shift at the M point upon monolayer of alkali metal dosing as a function of the difference between the work function of the sample and the electronegativity of dosed alkali metal.~$R^{2}=0.9715$ represents the coefficient of determination given by linear regression analysis. }
	\label{fig4}
\end{figure}

\section{4. Conclusion}

In conclusion, we have performed a ARPES study on the work function and alkali metal induced change in the electronic structure of three iron-based superconductors: FeSe, Ba(Fe$_{0.94}$Co$_{0.06}$)$_{2}$As$_{2}$ and NaFeAs. As previously proposed with an empirical relationship, our finding shows that the electronic structure change caused by monolayer of alkali metal dosing has a positive correlation with the difference between the work function of the sample and the electronegativity of dosed alkali metal. The present findings should offer experimental grounds for estimation of expected electronic structure change in ARPES studies using alkali metal dosing.

\section{Acknowledgements}
We are grateful to  Dr. W. S. Kyung, Dr. S. H. Cho, Y. S. Kim and B. J. Seok for their helpful discussions and useful comments. This work was conducted under the ISSP-CCES Collaborative Programme and was supported by the Institute for Basic Science in Republic of Korea (Grant Numbers IBS-R009-G2 and IBS-R009-Y2) and by JSPS KAKENHI 19K22140. The Advanced Light Source is supported by the Office of Basic Energy Sciences of the US DOE under Contract No. DE-AC02-05CH11231.

\end{document}